\documentclass[letterpaper, 10 pt, conference]{ieeeconf}  

\IEEEoverridecommandlockouts                              
\usepackage{ntheorem}
\usepackage[T1]{fontenc}

\usepackage{graphicx}      
\usepackage{newlfont}
\usepackage{dsfont}
\usepackage{amssymb,amsmath,amsfonts}
\usepackage{epstopdf}
\usepackage{empheq}

\newcommand\norm[1]{\left\lVert#1\right\rVert}

\newcommand{\Real}{{\mathds R}} 
\newcommand{\Nat}{{\mathds N}} 
{}
\newtheorem{corollary}{Corollary}{}
\newtheorem{proposition}{Proposition}{}
\newtheorem{theorem}{Theorem}{}
\newtheorem{remark}{Remark}{}
\newtheorem{lemma}{Lemma}{}

\title{\LARGE \bf
On Reachable Sets of Hidden CPS Sensor Attacks
}

\author{Carlos Murguia and Justin Ruths
\thanks{This work was supported by the National Research Foundation (NRF), Prime Minister's Office, Singapore, under its National Cybersecurity R\&D Programme (Award No. NRF2014NCR-NCR001-40) and administered by the National Cybersecurity R\&D Directorate.}
\thanks{C. Murguia is with the Engineering Systems and Design Pillar, Singapore University of Technology and Design. J. Ruths is with the Departments of Mechanical and Systems Engineering, University of Texas at Dallas. emails: murguia\_rendon@sutd.edu.sg \& jruths@utdallas.edu.}}

\begin{document}

\maketitle
\thispagestyle{empty}
\pagestyle{empty}

\begin{abstract}
For given system dynamics, observer structure, and observer-based fault/attack detection procedure, we provide mathematical tools -- in terms of Linear Matrix Inequalities (LMIs) -- for computing outer ellipsoidal bounds on the set of estimation errors that attacks can induce while maintaining the alarm rate of the detector equal to its attack-free false alarm rate. We refer to these sets to as \emph{hidden reachable sets}.\linebreak The obtained ellipsoidal bounds on hidden reachable sets quantify the attacker's potential impact when it is constrained to stay hidden from the detector. We provide tools for minimizing the volume of these ellipsoidal bounds (minimizing thus the reachable sets) by redesigning the observer gains. Simulation results are presented to illustrate the performance of our tools.
\end{abstract}

\section{Introduction}

There has recently been significant interest and work in the broad area of security of cyber-physical systems (CPS), see for example \cite{Cardenas}\nocite{Pasqualetti_1}\nocite{Mo_1}\nocite{Kwon}\nocite{Pappas}\nocite{Ahmed1}\nocite{Eric1}-\cite{Gupta2}. This topic investigates the properties of conventional control systems in the presence of adversarial disturbances. Control theory has shown great ability to robustly deal with disturbances and uncertainties \cite{Siep1}. However, adversarial attacks raise all-new issues due to the aggressive and strategic nature of the disturbances that attackers might inject into the system.

This paper focuses on attack detection and attack capabilities in CPSs. A majority of the work on attack detection leverages the established literature of fault detection \cite{Cardenas},\cite{Pasqualetti_1},\cite{Patton_1},\cite{Marios_Poly}. A fault detection approach uses an \emph{estimator} to forecast the evolution of the system dynamics. When the residual (the difference between what is measured and the estimation), or some function of the residual, is larger than a predetermined threshold, an alarm is raised. Arguably the most insidious attacks are those that occur without our knowledge. Fault detectors impose limits on the attacker, if the attacker aims to avoid being identified. Beyond retooling these existing methods for the new attack detection context, a fundamental question is: given a chosen fault detection approach, how does this method constrain the influence of an attacker? More specifically, what is an attacker able to accomplish when a system employs certain fault detection procedure?

Different methodologies exist for evaluating the impact of attacks. Most of the existing work uses some measure of state (or state estimate) deviation. In \cite{Pasqualetti_1}, the authors identify that if the attacker can take advantage of the zero dynamics of a (noise-free) input-output system, it can modify the system dynamics without reflecting its influence in the residual variables. This type of attacks are stealthy to any fault detector. A number of groups have studied the system response when the attacks are constrained by the detector. An important distinction between the collection of existing work -- and the work discussed here -- is the definition of how the attacker is constrained. We suggest the following terminology. While the term \textit{stealthy attack} is used very broadly, we suggest that this refer to the zero-dynamics case, as discussed in \cite{Pasqualetti_1}, because these attacks do not propagate to the residual. Some work has investigated the case of system response due to what we here call \textit{zero-alarm attacks}, i.e., attacks such that the detector threshold is never crossed \cite{Carlos_Justin1}-\nocite{Carlos_Justin2}\nocite{Jairo}\nocite{Cardenas_Sipolini}\cite{Guo2016}. Because real systems (with noise) always have a nonzero rate of false alarms raised by the detector, this attack model yields a relatively obvious attack signature because the alarms stop as soon as the attack starts. Other papers identify attacks that mimic the false alarm rate, thus making the alarm rate during the attack very close to the false alarm rate before the attack started \cite{Mo_3},\cite{Carlos_Justin3}. These attacks we call \textit{hidden attacks} because although they do change the distribution of the residual, these changes are hidden from the way the detector evaluates the distribution. A majority of this work uses state bounds or steady-state limits to quantify the impact that an attacker can have. The exceptions to this are \cite{Mo_3},\cite{Carlos_Justin3}, which quantify the reachable set of states and estimation errors when driven by the attack input.

This paper fuses several of these successful lines of research with a more strict interpretation of hidden attacks. The papers \cite{Mo_3},\cite{Carlos_Justin3} consider hidden attacks, however, they permit the alarm rate to change by a small value; the attacker capabilities that are derived are associated with this small deviation rather than the full scope of allowable attacks. Here, we fix the alarm rate exactly to study true hidden attacks (i.e., alarm rate exactly equal to the false alarm rate), and characterize the reachable sets on the estimation error dynamics associated with this broader definition of possible attack vectors. In this work, we characterize the \textit{hidden reachable sets} that the attacker can induce through manipulation of sensor data. Because in general, it is quite difficult to compute these sets exactly, for given system dynamics and attack detection scheme, we derive \emph{ellipsoidal bounds} on the hidden reachable sets using Linear Matrix Inequalities (LMIs)  \cite{BEFB:94}. Then, we provide synthesis tools for minimizing these bounds (minimizing thus the hidden reachable set) by properly redesigning the detectors.

This builds off of our previous work in \cite{Carlos_Justin3}. The strict interpretation of hidden attacks requires more direct handling of the effect of noise. To derive finite ellipsoidal bounds, we introduce the notion of $p$-probable reachable sets, which provides a nested set of ellipsoidal bounds based on the probability of the driving random sequences taking certain values. Because the derivation of the reachable set of states from the reachable set of estimation errors is captured in \cite{Carlos_Justin3} (for a class of observer-based output feedback controllers), and similar techniques can be used in this paper, we report here only on estimation error reachable sets. Note that the problem formulation in this paper, while seemingly similar, requires an entirely different characterization from \cite{Carlos_Justin3}.
\section{System Description \& Attack Detection}

\begin{figure}
  \centering
  \includegraphics[scale=.3]{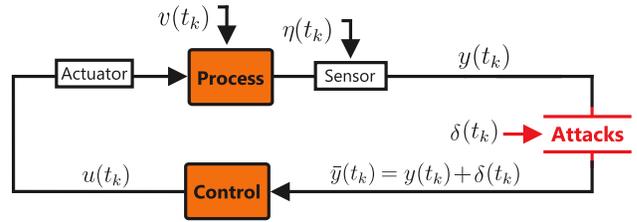}
  \caption{Cyber-physical system under attacks on the sensor measurements.}\label{Fig1}
\end{figure}

We study LTI stochastic systems of the form:
\begin{equation}
\left\{
\begin{array}{ll}
{x}(t_{k+1}) = Fx(t_k) + G u(t_k) + v(t_k),  \label{1} \\[1mm]
y(t_k) = Cx(t_k) + \eta(t_k),
\end{array}
\right.
\end{equation}
with sampling time-instants $t_k,k \in \Nat$,  state $x \in \Real^n$, measured output $y \in \Real^m$, control input $u \in \Real^l$, matrices $F$, $G$, and $C$ of appropriate dimensions, and i.i.d.  multivariate zero-mean Gaussian noises $v \in \Real^n$ and $\eta \in \Real^m$ with covariance matrices $R_{1} \in \Real^{n \times n}$, $R_1 \geq 0$ and $R_2 \in \Real^{m \times m}$, $R_2 \geq 0$, respectively. The initial state $x(t_1)$ is assumed to be a Gaussian random vector with covariance matrix $R_0 \in \Real^{n \times n}$, $R_0 \geq 0$. The processes $v(t_k)$, $k \in \Nat$ and $\eta(t_k)$, $k \in \Nat$ and the initial condition $x(t_1)$ are mutually independent. It is assumed that $(F,G)$ is stabilizable and $(F,C)$ is detectable. At the time-instants $t_k,k \in \Nat$, the output of the process $y(t_k)$ is sampled and transmitted over a communication network. The received output $\bar{y}(t_k)$ is used to compute control actions $u(t_k)$ which are sent back to the process, see Fig. \ref{Fig1}. The complete control-loop is assumed to be performed instantaneously, i.e., the sampling, transmission, and arrival time-instants are supposed to be equal. In this paper, we focus on attacks on sensor measurements. That is, in between transmission and reception of sensor data, an attacker may replace the signals coming from the sensors to the controller, see Fig. \ref{Fig1}. After each transmission and reception, the attacked output $\bar{y}$ takes the form:
\begin{equation}
  \bar{y}(t_k) := y(t_k) + \delta(t_k) = Cx(t_k) + \eta(t_k) + \delta(t_k), \label{3}
\end{equation}
where $\delta(t_k) \in \Real^m$ denotes \emph{additive sensor attacks}. Denote $x_k:=x(t_k)$, $u_k:= u(t_k)$, $v_k:=v(t_k)$, $\bar{y}_k:=\bar{y}(t_k)$, $\eta_k:=\eta(t_k)$, and $\delta_k:=\delta(t_k)$. Using this new notation, the attacked system is written as follows
\begin{equation}
\left\{
\begin{array}{ll}
{x}_{k+1} = F x_k + G u_k + v_k,\label{17} \\
\text{ \ \ }\hspace{.8mm}\bar{y}_k = C x_k + \eta_k + \delta_k.
\end{array}
\right.
\end{equation}

\subsection{Observer}

In order to estimate the state of the process, we use the following Luenberger observer \cite{Luen}
\begin{equation}
\hat{x}_{k+1} = F \hat{x}_k + Gu_k + L \big( \bar{y}_k - C\hat{x}_k   \big),  \label{19}
\end{equation}
with estimated state $\hat{x}_k \in \Real^n$, $\hat{x}_1 = E[x(t_1)]$, where $E[\hspace{.5mm}\cdot\hspace{.5mm}]$ denotes expectation, and observer gain matrix $L \in \Real^{n \times m}$. Define the estimation error $e_k:= x_k - \hat{x}_k$. Given the system dynamics (\ref{17}) and the observer (\ref{19}), the estimation error is governed by the following difference equation
\begin{align}
e_{k+1} = \big( F - L C \big) e_k - L \eta_k - L \delta_k + v_k. \label{20}\vspace{1mm}
\end{align}
The pair $(F,C)$ is detectable; hence, the observer gain $L$ can be selected such that $(F-LC)$ is Schur. Moreover, under detectability of $(F,C)$, \emph{if there are no attacks} (i.e., $\delta_k = \mathbf{0}$), where $\mathbf{0}$ denotes the zero matrix of appropriate dimensions, the covariance matrix $P_k:= E[e_ke_k^T]$ converges to steady state in the sense that $\lim_{k \rightarrow \infty} P_k = P$ exists, see \cite{Astrom}. For a given $L$ and $\delta_k=\mathbf{0}$, it can be verified that the asymptotic covariance matrix $P = \lim_{k \rightarrow \infty} P_k$ is given by the solution $P$ of the following Lyapunov equation:
\begin{align}
&(F-LC)P(F-LC)^T - P + R_1 + LR_2L^T = \mathbf{0}.\label{24}
\end{align}
It is assumed that the system has reached steady state before an attack occurs.

\subsection{Residuals and Hypothesis Testing}

In this manuscript, we characterize the effect that output injection attacks can induce in the system with being detected by \emph{fault detection techniques}. The main idea behind fault detection theory is the use of an estimator to forecast the evolution of the system. If the difference between what it is measured and the estimation is larger than expected, there may be a fault in or attack on the system. Although the notion of residuals and model-based detectors is now routine in the fault detection literature, the primary focus has been on detecting and isolating failures that have known signatures in the degradation of measurement quality, i.e., faults with specific structures. Now, in the context of an intelligent adversarial attacker for which there is no known attack signature, new challenges arise to understand the effect that an adaptive intruder can have on the system without being detected.
In this paper, we use the linear observer introduced in the previous section as our estimator. Define the \emph{residual sequence} $r_k, k\in \Nat$, as
\begin{align}
r_k := \bar{y}_k - C\hat{x}_k = Ce_k + \eta_k + \delta_k, \label{25}
\end{align}
which evolves according to the difference equation:
\begin{equation}
\left\{
\begin{array}{ll}
e_{k+1} = \big( F - LC \big) e_k - L\eta_k  - L \delta_k + v_k ,  \label{26} \\[.5mm]
\text{ \ \ } r_k = Ce_k + \eta_k + \delta_k.
\end{array}
\right.
\end{equation}
If there are no attacks, the steady state mean of $r_k$ is
\begin{equation}
E[r_{k+1}] = CE[e_{k+1}] + E[\eta_{k+1}] = \mathbf{0}_{m \times 1},  \label{27} \\
\end{equation}
and its asymptotic covariance matrix is given by
\begin{align}
\Sigma := E[r_{k+1}r_{k+1}^T] &= CPC^T + R_2.\label{28}
\end{align}
It is assumed that $\Sigma \in \Real^{m \times m}$ is positive definite. For this residual, we identify two hypotheses to be tested: $\mathcal{H}_0$ the \emph{normal mode} (no attacks) and $\mathcal{H}_1$ the \emph{faulty mode} (with faults/attacks). Then, we have
\begin{center}
\begin{tabular}{ c c c }
 $\mathcal{H}_0: \left\{
\begin{array}{ll}
E[r_k] = \mathbf{0}_{m \times 1},  \label{29} \\[.5mm]
E[r_kr_k^T] = \Sigma,
\end{array}
\right.$ & $\mathcal{H}_1: \left\{
\begin{array}{ll}
E[r_k] \neq \mathbf{0}_{m \times 1}, \text{or}  \label{30} \\[.5mm]
E[r_kr_k^T] \neq \Sigma,
\end{array}
\right.$
\end{tabular}
\end{center}
where $\mathbf{0}_{m \times 1}$ denotes an $m$-dimensional vector composed of zeros only. In this manuscript, we use the chi-squared procedure for examining the residual and subsequently detecting attacks.

\subsection{Distance Measure and Chi-squared Procedure}

The input to any detection procedure is a \emph{distance measure} $z_k \in \Real$, $k \in \Nat$, i.e., a measure of how deviated the estimator is from the sensor measurements. We employ distance measures any time we test to distinguish between $\mathcal{H}_0$ and $\mathcal{H}_1$. The chi-squared test uses a quadratic form on the residual as distance measure to test for substantial variations in mean and variance of the error between the measured output and the estimate. Consider the residual sequence $r_{k}$, \eqref{26}, and its covariance matrix $\Sigma$, \eqref{28}. The chi-squared procedure is defined as follows.

\noindent\rule{\hsize}{1pt}\vspace{.2mm}
\textbf{Chi-squared procedure:}
\begin{equation}\label{baddata}
\text{If \ } z_k := r_{k}^T \Sigma^{-1} r_k  > \alpha, \hspace{2mm} \tilde{k} = k.
\end{equation}
\textbf{Design parameter:} threshold $\alpha \in \Real_{>0}$.\\
\textbf{Output:} alarm time(s) $\tilde{k}$.\\
\vspace{.2mm}\noindent\rule{\hsize}{1pt}\vspace{1mm}
Thus, the procedure is designed so that alarms are triggered if $z_{k}$ exceeds the threshold $\alpha$. The normalization by $\Sigma^{-1}$ makes setting the value of the threshold $\alpha$ system independent. This quadratic expression leads to a sum of the squares of $m$ normally distributed random variables which implies that the distance measure $z_k$ follows a chi-squared distribution with $m$ degrees of freedom, see, e.g., \cite{Ross} for details.


\subsection{False Alarms}\label{False_alarma}

The occurrence of an alarm in the chi-squared procedure when there are no attacks to the CPS is referred to as a false alarm. The threshold $\alpha$ must be selected to fulfill a \emph{desired false alarm rate} $\mathcal{A}^*$. Let $\mathcal{A} \in [0,1]$ denote the \emph{false alarm rate} of the chi-squared procedure defined as {the expected proportion of observations which are false alarms}, i.e., $\mathcal{A}:=\text{pr}[z_k \geq \alpha]$, where $\text{pr}[\cdot]$ denotes probability, see \cite{Dobben} and \cite{Adams}. 

\begin{proposition}\label{prop3}{\emph{\cite{Carlos_Justin2}}}. Assume that there are no attacks on the system and consider the chi-squared procedure \eqref{baddata} with residual $r_{k} \sim \mathcal{N}(\mathbf{0},\Sigma)$ and threshold $\alpha \in \Real_{>0}$. Let  $\alpha = \alpha^* := 2 \text{\emph{P}}^{-1}(\frac{m}{2},1-\mathcal{A}^*)$, where $\emph{\text{P}}^{-1}(\cdot,\cdot)$ denotes the inverse regularized lower incomplete gamma function (see \emph{\cite{Ross}}), then $\mathcal{A} = \mathcal{A}^*$.
\end{proposition}

%

\section{Hidden Reachable Sets}

In this section, we provide tools for \emph{quantifying} (for given $L$) and \emph{minimizing} (by selecting $L$) the impact of the attack sequence $\delta_{k}$ on the estimation error $e_k$ when the chi-squared procedure is used for attack detection. To quantify the effect of attacks, we need to introduce some measure of impact. However, because malicious adversaries may launch any arbitrary attack, we need a measure which can capture all possible trajectories that the attacker can induce in the estimation error dynamics, given how it accesses the dynamics (i.e., through residual variables by tampering with sensor measurements). We propose to use the \emph{reachable set} of the attack sequence $\delta_k$ as our measure of impact. We are interested in \emph{attacks that do not change the false alarm rate} of the detector $\mathcal{A}$, i.e., $\bar{\mathcal{A}} = \mathcal{A}$, where $\bar{\mathcal{A}}$ denotes the alarm rate under the attacker's action. This class of attacks is what we refer to as \emph{hidden attacks} and the trajectories that hidden attacks can induce in the system are referred to as \emph{hidden reachable sets}. In this section, we provide tools based on Linear Matrix Inequalities (LMIs) for computing outer ellipsoidal bounds on the hidden reachable sets induced by the attack sequence $\delta_k$ given the system dynamics, the chi-squared procedure, the noise, and the false alarm rate $\mathcal{A}$.

\subsection{Attack Model and Hidden Reachable Sets}

We assume that the attacker has perfect knowledge of the system dynamics, the observer,  measurements, and detection procedure (chi-squared). It is further assumed that all the sensors can be compromised by the attacker\- at each time step (the case where not all the sensors are attacked is left as future work). By considering this strong, worst-case attacker, we are able to construct an upper bound on the abilities of the attacker.
Consider the estimation error dynamics (\ref{26}), the residual sequence $r_k = Ce_k + \eta_k + \delta_k$, and the distance measure
\begin{equation}\label{DMA}
z_k = ||\Sigma^{-\frac{1}{2}}r_{k}||^2 = ||\Sigma^{-\frac{1}{2}}(Ce_{k} + \eta_k + \delta_k)||^2,
\end{equation}
where $\Sigma^{-\frac{1}{2}}$ denotes the symmetric squared root matrix of $\Sigma^{-1}$. The set of feasible attack sequences that the opponent can launch while satisfying $\bar{\mathcal{A}} = \mathcal{A}$ can be written as the following constrained control problem on $\delta_k$:
\begin{equation}\label{constrained_control}
\resizebox{.43 \textwidth}{!}
{
$
\left\{ \delta_k \in \Real^m \left|
\begin{array}{ll}
e_k \text{ satisfies \eqref{26}, and}\\[1mm]
\text{pr}[\hspace{.01mm}||\Sigma^{-\frac{1}{2}}(Ce_{k} + \eta_k + \delta_k)||^2 > \alpha \hspace{1mm}] = \mathcal{A},
\end{array}
\right. \right\},
$
}
\end{equation}
for $k \in \Nat$.
We are interested in the error trajectories that the attacker can induce in the system restricted to satisfy \eqref{constrained_control}. Note that, as long as $\bar{\mathcal{A}}=\mathcal{A}$, the attacker may induce any arbitrary random sequence $\delta_k$. This and the fact that $v_k$ and $\eta_k$ are Gaussian (thus having infinite support) imply that deterministic reachable sets induced by $\delta_k$ and the noise sequences are generally unbounded. To overcome this obstacle, we introduce the notion of \emph{$p$-probable hidden reachable sets} $\mathcal{R}_{\alpha}^p$. Define $\zeta_k := \Sigma^{-\frac{1}{2}}(Ce_{k} + \eta_k + \delta_k)$ and note that the estimation error dynamics (\ref{26}) can be written in terms of $\zeta_k$ as:
\begin{equation}\label{75prime}
\left\{
\begin{array}{lll}
\hspace{.3mm} e_{k+1} = F e_k - L\Sigma^{\frac{1}{2}} \zeta_k + v_k, \\
\hspace{4mm} \zeta_k = \Sigma^{-\frac{1}{2}}(Ce_{k} + \eta_k + \delta_k).
\end{array} \right.
\end{equation}
For given false alarm rate $\mathcal{A}$ and probability $p \in (0,1)$, \emph{the $p$-probable hidden reachable set} of the attack sequence $\delta_k$ in \eqref{75prime}, $\mathcal{R}_{\alpha}^p$, is defined as the set of $e_k \in \Real^n$, $k\in \Nat$ that can be reached from the origin $e_1 = \mathbf{0}$ due to the the attacker's action $\delta_k$ restricted to satisfy $\bar{\mathcal{A}}=\mathcal{A}$ and
\begin{equation}\label{restriction}
\text{pr}[\hspace{.1mm}||\zeta_k||^2 \leq \bar{\zeta}_p \hspace{.1mm}] =
\text{pr$[\norm{v_k}^2 \leq \bar{v}_p]=p,$}
\end{equation}
for some constants $\bar{\zeta}_p,\bar{v}_p \in \Real_{>0}$, i.e.,
\begin{equation}\label{constrained_control4}
\resizebox{.43 \textwidth}{!}
{
$
\mathcal{R}_{\alpha}^p:=\left\{e_{k} \in \Real^n \left|
\begin{array}{ll}
e_{1}=\mathbf{0},\\[.1mm] 
e_{k},\delta_k,v_k \hspace{1mm} \text{satisfy} \hspace{1mm}  \text{\eqref{constrained_control}-\eqref{restriction}},
\end{array}
\right. \right\}.
$
}
\end{equation}
By restricting the probabilities in \eqref{restriction}, we are delimiting the support of the attack and noise sequences to compact sets. Then, the $p$-probable hidden reachable sets correspond to the trajectories of the system when the driving random sequences are restricted to satisfy $\bar{\mathcal{A}} = \mathcal{A}$ and \eqref{restriction}. For delimited $v_k$ and $\delta_k$, we can characterize reachable sets using deterministic tools. In general, it is analytically intractable to compute $\mathcal{R}_{\alpha}^p$ exactly. Instead, using LMIs, for some positive definite matrix $\mathcal{P}_{\alpha}^p \in \Real^{n \times n}$, we derive outer ellipsoidal bounds of the form $\mathcal{E}_{\alpha}^p := \{ e_{k} \in \Real^{n} | e_{k}^T \mathcal{P}_{\alpha}^p e_{k} \leq 1 \}$ containing $\mathcal{R}_{\alpha}^p$.

\begin{remark}
Note, from \eqref{75prime}, that if for some $k=k^*$, $e_{k^*} \neq 0$ and $\rho[F] > 1$, where $\rho[\cdot]$ denotes spectral radius, then $||e_{k}||$ diverges to infinity as $k \rightarrow \infty$ for any non-stabilizing $\zeta_k$. That is, $\mathcal{R}_{\alpha}^p$ is unbounded if the system is open-loop unstable. If $\rho[F] \leq 1$, then $||e_{k}||$ may or may not diverge to infinity depending on algebraic and geometric multiplicities of the eigenvalues with unit modulus of $F$ (a known fact from stability of LTI systems), see \emph{\cite{Astrom}} for details.
\end{remark}

Given Remark 1, in what follows, we consider open-loop stable systems ($\rho[F]<1$). The following result is used to compute the ellipsoidal bounds $\mathcal{E}_{\alpha}^p$.

\begin{lemma} \label{Lemma1} \emph{\cite{Rechable_set_1}}
Let $\xi_k \in \Real^n$, $\xi_1 = \mathbf{0}$, $V_k := \xi_k^T \mathcal{P} \xi_k$, for some positive definite matrix $\mathcal{P} \in \Real^{n \times n}$, and $\omega_k^T \omega_k \leq \bar{\omega}$, $\bar{\omega} \in \Real_{>0}$. If there exists a constant $b \in (0,1)$ such that
\begin{equation}\label{Lemma1_Rechable_set_1}
V_{k+1} - bV_k - \frac{1-b}{\bar{\omega}} \omega_k^T \omega_k \leq 0, \forall \hspace{1mm} k \in \Nat,
\end{equation}
then, $V_k = \xi_k^T \mathcal{P} \xi_k \leq 1$.
\end{lemma}

\subsection{Case 1: $p \in [0,1-\mathcal{A}]$}

Because the attack sequence is restricted to satisfy \eqref{constrained_control}, we start computing the ellipsoidal bounds corresponding to $p = 1-\mathcal{A}$, i.e., $\mathcal{E}_{\alpha}^{1-\mathcal{A}}$. It is easy to verify using Lemma 1 that $\mathcal{E}_{\alpha}^p \subseteq \mathcal{E}_{\alpha}^{1-\mathcal{A}}$ for $p \in [0,1-\mathcal{A}]$ because $\bar{\zeta}_p \leq \bar{\zeta}_{1-\mathcal{A}} = \alpha$ and $\bar{v}_p \leq \bar{v}_{1-\mathcal{A}}$ in \eqref{restriction}; i.e., all $p$-probable ellipsoidal bounds for $p\in[0,1-\mathcal{A}]$ lie within the $1-\mathcal{A}$-probable ellipsoidal bound. It follows that $\mathcal{R}_e^p \subseteq \mathcal{E}_{\alpha}^{1-\mathcal{A}}$ for $p \in [0,1-\mathcal{A}]$, i.e., for $p$ in this interval, we only need to compute the ellipsoidal bound corresponding to $p = 1-\mathcal{A}$. Characterizing $p$-probable sets for small $p$ values is of little interest because they do not provide a informative bound on system trajectories (since the smaller $p$ is, the more trajectories lie outside the $p$-probable ellipsoidal bound). We work with the data available in this setting, namely the number of alarms raised by the detector, to bound the most informative $p=1-\mathcal{A}$ probable reachable set; in Case 2, we extend these results for larger $p$ values.

\begin{theorem}\label{Theo2}
For given system matrix $F$, observer gain $L$, residual covariance matrix $\Sigma$, and false alarm rate $\mathcal{A}$, consider the set $\mathcal{R}_{\alpha}^{1-\mathcal{A}}$ in  \eqref{constrained_control4}. If there exists a positive definite matrix $\mathcal{P}  \in \Real^{n \times n}$ and $b \in (0,1)$ satisfying the following matrix inequality:
\begin{align}\begingroup
\renewcommand*{\arraycolsep}{1.5pt}
\begin{array}{ll}\label{LMI_1}
\begin{bmatrix}
b\mathcal{P} & F^T \mathcal{P}  & \mathbf{0} & \mathbf{0} & \mathbf{0} & \mathbf{0}\\
\mathcal{P} F & \mathcal{P} & \mathcal{P} & -\mathcal{P} L \Sigma^{\frac{1}{2}} & \mathbf{0} & \mathbf{0}\\
\mathbf{0} & \mathcal{P} & \tfrac{1-b}{\bar{\omega}}I & \mathbf{0} & \mathbf{0} & \mathbf{0}\\
\mathbf{0} & -\Sigma^{\frac{1}{2}}L^T \mathcal{P} & \mathbf{0} & \tfrac{1-b}{\bar{\omega}}I & \mathbf{0} & \mathbf{0}\\
\mathbf{0} & \mathbf{0} & \mathbf{0} & \mathbf{0} & I & \mathbf{0}\\
\mathbf{0} & \mathbf{0} & \mathbf{0} & \mathbf{0} & \mathbf{0} & I
\end{bmatrix} &\geq \mathbf{0};
\end{array}\endgroup
\end{align}
for $\bar{\omega} = \alpha + \bar{v}_{1-\mathcal{A}}$; then, $\mathcal{R}_{\alpha}^{1-\mathcal{A}} \subseteq \mathcal{E}_{\alpha}^{1-\mathcal{A}}$ with $\mathcal{P}_{\alpha}^{1-\mathcal{A}} = \mathcal{P}$, i.e., the $(1-\mathcal{A})$-probable hidden reachable set is contained in the ellipsoid $\mathcal{E}_{\alpha}^{1-\mathcal{A}}=\{ e_{k} \in \Real^{n} | e_{k}^T \mathcal{P}_{\alpha}^{1-\mathcal{A}} e_{k} \leq 1 \}$.
\end{theorem}
\emph{\textbf{Proof}}: For a positive definite matrix $\mathcal{P}  \in \Real^{n \times n}$, consider the function $V_k := e_{k}^T \mathcal{P} e_{k}$, then, from \eqref{constrained_control4}, inequa\-lity \eqref{Lemma1_Rechable_set_1} takes the form:
\begin{equation*}
\resizebox{.485 \textwidth}{!}
{
$
\begingroup
\renewcommand*{\arraycolsep}{.5pt}
\begin{array}{ll}
=-\vartheta_k^T
\begin{bmatrix} b\mathcal{P} - F^T \mathcal{P} F & F^T \mathcal{P} L \Sigma^{\frac{1}{2}} & -F^T\mathcal{P} \\ \Sigma^{\frac{1}{2}}L^T \mathcal{P} F & \tfrac{1-b}{\bar{\omega}}I -\Sigma^{\frac{1}{2}}L^T \mathcal{P} L\Sigma^{\frac{1}{2}} & \Sigma^{\frac{1}{2}}L^T \mathcal{P}\\
-\mathcal{P}F & \mathcal{P}L\Sigma^{\frac{1}{2}} & \tfrac{1-b}{\bar{\omega}}I - \mathcal{P}
 \end{bmatrix}
\vartheta_k\\[5mm]
=: - \vartheta_k^T \mathcal{Q}_e \vartheta_k  \leq 0,
\end{array}\endgroup
$
}
\end{equation*}
where $\vartheta := (e_{k}^T,\zeta_k^T,v_k^T)^T$. The above inequality is satisfied if and only if $\mathcal{Q}_e \geq \mathbf{0}$. Matrix $\mathcal{Q}_e$ can be written as the Schur complement of a higher dimensional matrix $\mathcal{Q}_e^{\prime}$; hence, $\mathcal{Q}_e \geq \mathbf{0} \leftrightarrow \mathcal{Q}_e^{\prime} \geq \mathbf{0}$, i.e.,
\begin{equation} \label{LMIProof1}
\resizebox{.4 \textwidth}{!}
{
$
\begingroup
\renewcommand*{\arraycolsep}{.5pt}
\begin{array}{ll}
\mathcal{Q}_e \geq \mathbf{0} \leftrightarrow\\
\mathcal{Q}_e^{\prime} := \begin{bmatrix}
b\mathcal{P} & \mathbf{0} & \mathbf{0} & F^T \mathcal{P} & \mathbf{0} & \mathbf{0} \\ \mathbf{0} & \tfrac{1-b}{\bar{\omega}}I & \mathbf{0} & -\Sigma^{\frac{1}{2}}L^T\mathcal{P} & \mathbf{0} & \mathbf{0} \\ \mathbf{0} & \mathbf{0} & \tfrac{1-b}{\bar{\omega}}I & \mathcal{P} & \mathbf{0} & \mathbf{0}\\
\mathcal{P}F & -\mathcal{P}L\Sigma^{\frac{1}{2}} & \mathcal{P} & \mathcal{P} & \mathbf{0} & \mathbf{0} \\ \mathbf{0} & \mathbf{0} & \mathbf{0} & \mathbf{0} & I & \mathbf{0}\\ \mathbf{0} & \mathbf{0} & \mathbf{0} & \mathbf{0} & \mathbf{0} & I
\end{bmatrix} \geq \mathbf{0}.
\end{array}
\endgroup
$
}
\end{equation}
Finally, inequality \eqref{LMI_1} follows from \eqref{LMIProof1}  by a simple reordering of rows and columns.
The result follows now from Lemma 1 by taking $\mathcal{P}_{\alpha}^{1-\mathcal{A}}=\mathcal{P}$ and $\bar{\omega}=\alpha + \bar{v}_{1-\mathcal{A}}$. \hfill $\blacksquare$

The result in Theorem 1 provides a tool for computing\- ellipsoidal bounds on $\mathcal{R}_{\alpha}^{1-\mathcal{A}}$. To make the bounds most useful, we next construct ellipsoids with minimal volume, i.e., the tightest possible ellipsoid bounding $\mathcal{R}_{\alpha}^{1-\mathcal{A}}$. In this case, we have to minimize $\det \mathcal{P}^{-1}$ subject to \eqref{LMI_1} (because $\det \mathcal{P}^{-1}$ is proportional to the volume of $e_{k}^T \mathcal{P} e_{k} = 1$). This is formally stated in the following corollary of Theorem 1, see \cite{BEFB:94} for further details.

\begin{corollary}\label{remark1}
For given matrices $(F,L,\Sigma)$, false alarm rate $\mathcal{A}$, and $b \in (0,1)$, the solution $\mathcal{P}$ of the following convex optimization:
\begin{align}
\left\{ \begin{array}{ll}\label{LMI_2}
\min_{\mathcal{P}} \hspace{2mm} -\log\det \mathcal{P}, \\[1mm]
\text{s.t. } \mathcal{P}>0 \text{ and } \eqref{LMI_1},
\end{array} \right.
\end{align}
for $\bar{\omega} = \alpha + \bar{v}_{1-\mathcal{A}}$, minimizes the volume of the ellipsoid $\mathcal{E}_{\alpha}^{1-\mathcal{A}}$ (with $\mathcal{P}_{\alpha}^{1-\mathcal{A}} = \mathcal{P}$) bounding $\mathcal{R}_{\alpha}^{1-\mathcal{A}}$.
\end{corollary}

See \cite{YALMIP} for an example of how to solve (\ref{LMI_2}) using YALMIP.

As we now move toward redesigning $L$ to minimize the ellipsoids, we note that as $||L||\rightarrow 0$, the volume of $\mathcal{E}_{\alpha}^{1-\mathcal{A}}$ goes to zero because the attack-dependent term in \eqref{75prime}, $L\Sigma^{\frac{1}{2}}\zeta_k$, vanishes. In other words, without any other considered criteria, the observer gain leading to the minimum volume ellipsoid is trivially given by $L=\mathbf{0}$. While this is effective at eliminating the impact of the attacker, it implies that we discard the observer altogether and, therefore, forfeit any ability to build a reliable estimate of the system state. If we impose a performance criteria that the observer must satisfy in the attack-free case (e.g., convergence speed, noise-output gain, and minimum asymptotic variance), it has to be added into the minimization problem \eqref{LMI_2} so as to minimize the volume of $\mathcal{E}_{\alpha}^{1-\mathcal{A}}$ while still achieving the observer performance in the attack-free case. For completeness, in the following proposition, we provide an LMI criteria for ensuring that the $H_{\infty}$ gain from the noise to the residual $r_k$ in \eqref{26} is less than or equal to some $\gamma \in \Real_{>0}$. Then, using this criteria and Theorem \ref{Theo2}, we provide a synthesis tool for minimizing the volume of $\mathcal{E}_{\alpha}^{1-\mathcal{A}}$ while ensuring a desired $H_{\infty}$ performance in the attack-free case.

\begin{proposition} \label{prop:designL}
For given matrices $(F,C,L)$, if there exist a positive definite matrix $\mathcal{P}  \in \Real^{n \times n}$ and constant $\gamma \in \Real_{>0}$ satisfying the following matrix inequality:
\begin{equation}\label{LMI_Hinf}
\resizebox{.4325 \textwidth}{!}
{
$
\begingroup
\renewcommand*{\arraycolsep}{2pt}
\begin{array}{ll}
\begin{bmatrix} \mathcal{P} & \mathbf{0} & \mathbf{0} & (F - LC)^T\mathcal{P}  & C^T\\
\mathbf{0} & \gamma^2I & \mathbf{0} & -L^T\mathcal{P} & I\\
\mathbf{0} & \mathbf{0} & \gamma^2I & \mathcal{P} & \mathbf{0}\\
\mathcal{P}(F-LC) & -\mathcal{P}L & \mathcal{P} & \mathcal{P} & \mathbf{0}\\
C & I & \mathbf{0} & \mathbf{0} & I
\end{bmatrix} \geq \mathbf{0},
\end{array}\endgroup
$
}
\end{equation}
then, the $H_{\infty}$ gain from the noise $\nu_k := (\eta_k^T,v_k^T)^T$ to the residual $r_k = Ce_{k} + \eta_k$ of the estimation error dynamics \eqref{26} is less than or equal to $\gamma$.
\end{proposition}
The proof of Proposition \ref{prop:designL} is omitted here due to the page limit. However, this is a standard result and details about the proof can be found in, e.g., \cite{Siep1} and references therein. In the following corollary of Theorem \ref{Theo2} and Proposition \ref{prop:designL},\linebreak we formulate the optimization problem for designing the observer gain $L$ such that the volume of the ellipsoid $\mathcal{E}_{\alpha}^{1-\mathcal{A}}$ is minimized and a desired $H_{\infty}$ performance is achieved in the attack-free case.

\begin{corollary}\label{remark2}
For given system matrices $(F,C)$, residual \linebreak covariance matrix $\Sigma$, false alarm rate $\mathcal{A}$, $b \in (0,1)$, and $\gamma \in \Real_{>0}$, if there exist matrices $\mathcal{P} \in \Real^{n \times n}$ and $M \in \Real^{n \times m}$\linebreak solution to the following convex optimization:
\begin{align}\begingroup
\renewcommand*{\arraycolsep}{1.0pt}
\left\{ \begin{array}{ll}\label{LMI_33}
\min_{\mathcal{P},M} \hspace{2mm} -\log\det \mathcal{P}, \\[2mm]
\text{s.t. } \mathcal{P}>0, \small
\begin{bmatrix}
b\mathcal{P} & F^T \mathcal{P}  & \mathbf{0} & \mathbf{0} & \mathbf{0} & \mathbf{0}\\
\mathcal{P} F & \mathcal{P} & \mathcal{P} & -M \Sigma^{\frac{1}{2}} & \mathbf{0} & \mathbf{0}\\
\mathbf{0} & \mathcal{P} & \tfrac{1-b}{\bar{\omega}}I & \mathbf{0} & \mathbf{0} & \mathbf{0}\\
\mathbf{0} & -\Sigma^{\frac{1}{2}}M^T & \mathbf{0} & \tfrac{1-b}{\bar{\omega}}I & \mathbf{0} & \mathbf{0}\\
\mathbf{0} & \mathbf{0} & \mathbf{0} & \mathbf{0} & I & \mathbf{0}\\
\mathbf{0} & \mathbf{0} & \mathbf{0} & \mathbf{0} & \mathbf{0} & I
\end{bmatrix} \geq \mathbf{0}, \normalsize \text{ and}\\[11mm]
\small \begin{bmatrix} \mathcal{P} & \mathbf{0} & \mathbf{0} & F^T\mathcal{P} - C^TM^T  & C^T\\
\mathbf{0} & \gamma^2I & \mathbf{0} & -M^T & I\\
\mathbf{0} & \mathbf{0} & \gamma^2I & \mathcal{P} & \mathbf{0}\\
\mathcal{P}F-MC & -M\text{ } & \mathcal{P} & \mathcal{P} & \mathbf{0}\\
C & I & \mathbf{0} & \mathbf{0} & I
\end{bmatrix} \geq \mathbf{0},\normalsize
\end{array} \right. \endgroup
\end{align}
for $\bar{\omega} = \alpha + \bar{v}_{1-\mathcal{A}}$; then, the observer gain $L= \mathcal{P}^{-1} M$ minimizes the volume of the ellipsoid $\mathcal{E}_{\alpha}^{1-\mathcal{A}}$ (with $\mathcal{P}_{\alpha}^{1-\mathcal{A}} = \mathcal{P}$) bounding $\mathcal{R}_{\alpha}^{1-\mathcal{A}}$ and guarantees that the $H_{\infty}$ gain from the noise $\nu_k = (\eta_k^T,v_k^T)^T$ to the residual $r_k$ of \eqref{26} is less than or equal to $\gamma$ in the attack-free case.
\end{corollary}
\emph{\textbf{Proof}}: This follows from Theorem \ref{Theo2}, Proposition \ref{prop:designL}, and the linearizing change of variables $M=\mathcal{P} L$.  \hfill $\blacksquare$

\subsection{Case 2: $p \in (1-\mathcal{A},1]$}

Note that, for $p \in (1-\mathcal{A},1]$, $\bar{\zeta}_p > \bar{\zeta}_{1-\mathcal{A}} = \alpha$ according to \eqref{restriction}. Then, we can write $\bar{\zeta}_p = \alpha + \epsilon_p$ and $\text{pr}[\norm{\zeta_k}^2 \leq \alpha + \epsilon_p] = 1 - \mathcal{A} + a_p$, for some $\epsilon_p \in (0,\infty)$ and $a_p \in (0,\mathcal{A}]$. To be able to compute ellipsoidal bounds, the constant $\epsilon_p$ corresponding to a given probability $1-\mathcal{A}+a_p$ is required. If $\epsilon_p$ is available, we can restrict $\zeta_k$ to compact sets as in Case 1.\linebreak Note, however, that the distribution of the attack sequence $\delta_k$ (and thus the one of $\zeta_k$) is generally unknown. Actually, the attacker may induce any arbitrary (and possibly) non-stationary random sequence $\zeta_k$ in \eqref{75prime} as long as $\bar{\mathcal{A}}=\mathcal{A}$. Nevertheless, we can obtain bounds on $\epsilon_p$ using Markov's inequality \cite{Ross} to link the statistical properties of $\zeta_k$ with $\epsilon_p$. This is stated in the following proposition.

\begin{proposition}\label{prop:case2}
Denote $\mathcal{M}_k:=E[\zeta_k \zeta_k^T]$ and $\mu_k:=E[\zeta_k]$. For given false alarm rate $\mathcal{A}$, probability $p=1-\mathcal{A}+a_p$, and $a_p \in (0,\mathcal{A})$, the following is satisfied:
\begin{subequations}\label{lower_bound}
\begin{empheq}[left=\empheqlbrace]{align}
&\text{\emph{pr}}[\norm{\zeta_k}^2 \leq \alpha + \epsilon_p] \in [1-\mathcal{A}+a_p,1],\label{lower_bounda}\\[1mm]
&\text{\emph{for all} }\epsilon_p \geq \underline{\epsilon}_p:= \dfrac{\text{\emph{tr}}[\mathcal{M}_k] + \mu_k^T \mu_k}{\mathcal{A}-a_p}-\alpha. \label{lower_boundb}
\end{empheq}
\end{subequations}
\end{proposition}
\emph{\textbf{Proof}}: The probability $\text{pr}[\norm{\zeta_k}^2 \leq \alpha + \epsilon_p]$ can be written as  $\text{pr}[\norm{\zeta_k}^2 \leq \alpha + \epsilon_p] = 1 - \text{pr}[\norm{\zeta_k}^2 > \alpha + \epsilon_p]$.
Then, using Markov's inequality \cite{Ross}, we can write the following
\[
\text{pr}[\norm{\zeta_k}^2 > \alpha + \epsilon_p] = 1-\text{pr}[\norm{\zeta_k}^2 \leq \alpha + \epsilon_p]\leq \cfrac{E[\norm{\zeta_k}^2]}{\alpha + \epsilon_p}.
\]
Therefore, if $\epsilon_p$ satisfies $E[\norm{\zeta_k}^2]/(\alpha + \epsilon_p) \leq \mathcal{A} - a_p$, then $\text{pr}[\norm{\zeta_k}^2 > \alpha + \epsilon_p] \leq \mathcal{A} - a_p$ and hence $\text{pr}[\norm{\zeta_k}^2 \leq \alpha + \epsilon_p] \in [1-\mathcal{A} + a_p,1]$. The expectation of the quadratic form $\zeta_k^T\zeta_k$ is given by $E[\norm{\zeta_k}^2]=E[\zeta_k^T \zeta_k] = \text{tr}[\mathcal{M}_k] + \mu_k^T\mu_k$ \cite{Ross}; then, $E[\norm{\zeta_k}^2]/(\alpha + \epsilon_p) \leq \mathcal{A} - a_p$ is satisfied for all $\epsilon_p \geq \underline{\epsilon}_p$ with $\underline{\epsilon}_p$ as defined in \eqref{lower_boundb}, and the assertion follows. \hfill $\blacksquare$

Using Proposition 3, for given false alarm rate $\mathcal{A}$ and probability $p = 1-\mathcal{A}+a_p \in (1-\mathcal{A},1]$, $a_p \in (0,\mathcal{A})$, we can characterize $p$-probable hidden reachable sets, $\tilde{\mathcal{R}}_{\alpha}^p$, by using the lower bounds on $\text{pr}[\norm{\zeta_k}^2 \leq \alpha + \epsilon_p]$ and $\epsilon_p$ in \eqref{lower_bound}. Specifically, for $p>1-\mathcal{A}$, the set $\tilde{\mathcal{R}}_{\alpha}^p$ of the sequence $\delta_k$ is defined as the set of $e_k \in \Real^n$ that can be reached from $e_1 = \mathbf{0}$ restricted to satisfy $\bar{\mathcal{A}}=\mathcal{A}$ and
\begin{equation}\label{restriction2}
\resizebox{.4325 \textwidth}{!}
{
$
\left\{
\begin{array}{ll}
\text{pr}[\norm{v_k}^2 \leq \bar{v}_p]=1-\mathcal{A} + a_p \text{ and}\\
\epsilon_p = \underline{\epsilon}_p \rightarrow \text{pr}[\norm{\zeta_k}^2 \leq \alpha + \epsilon_p] \in [1-\mathcal{A} + a_p,1],
\end{array}
\right.
$
}
\end{equation}
for some constant $\bar{v}_p \in \Real_{>0}$ and $\underline{\epsilon}_p$ as defined in (\ref{restriction2}), i.e.,
\begin{equation}\label{constrained_control6}
\resizebox{.4325 \textwidth}{!}
{
$
\tilde{\mathcal{R}}_{\alpha}^p:=\left\{e_{k} \in \Real^n \left|
\begin{array}{llll}
e_{1}=\mathbf{0},\\
\text{$e_{k},\zeta_k,v_k$, satisfy} \hspace{1mm}  \text{\eqref{constrained_control}-\eqref{75prime},\eqref{restriction2},}
\end{array}
\right. \right\}.
$
}
\end{equation}

\begin{remark}
For a $p$-probable reachable set, we select $a_p$ such that $p=1-\mathcal{A}+a_p$, then determine $\epsilon_p$ using \eqref{lower_boundb}. Note that, because the attacker can induce an attack sequence with arbitrarily large covariance $\mathcal{M}_k$ and mean $\mu_k$, the lower bound on $\epsilon_p$, $\underline{\epsilon}_p$, in \eqref{lower_boundb} can be made arbitrarily large for any $a_p$. Therefore, if $\delta_k$ (and thus $\zeta_k$) is only restricted to satisfy $\bar{\mathcal{A}}=\mathcal{A}$, the opponent can induce arbitrarily large reachable sets $\tilde{\mathcal{R}}_{\alpha}^p$.
\end{remark}

\begin{figure}[t]\label{Fig2}
  \centering
  \includegraphics[scale=.31]{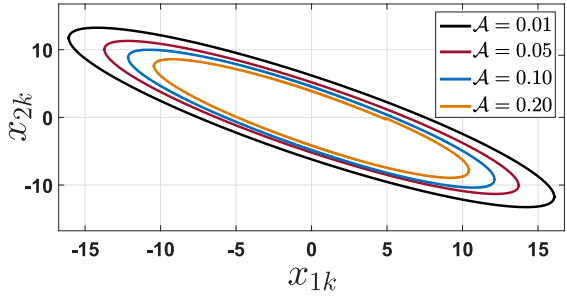}
  \caption{Ellipsoid $\mathcal{E}_{\alpha}^{1-\mathcal{A}}$ for different values of false alarm rate $\mathcal{A}$.}\label{Fig2}
\end{figure}

Remark 2 implies that if we only monitor the alarms raised by the detector, the attacker can inject arbitrarily large signals in the residual sequence $r_k$ without changing the alarm rate. Consequently, the sets $\tilde{\mathcal{R}}_{\alpha}^p$ can be made arbitrarily large for arbitrarily small $a_p$. If we place additional assumptions on the attacker, namely that the mean and covariance of the attack sequence $\zeta_k$ are finite, the reachable sets will be bounded by Proposition \ref{prop:case2}. In particular, if we assume the attacker maintains the mean and covariance of the attack-free scenario, i.e., $E[\zeta_k]=\mu_k=\mathbf{0}$ and $E[\zeta_k\zeta_k^T]=\mathcal{M}_k=I_m$, then $\underline{\epsilon}_p = \frac{m}{\mathcal{A} - a_p} -\alpha$. Hence, if in addition to imposing $\bar{\mathcal{A}} = \mathcal{A}$, the attack is restricted to keep the statistical properties of $\zeta_k$ in the attack-free case, i.e., $\mu_k=\mathbf{0}$ and $\mathcal{M}_k=I_m$, the reachable sets $\tilde{\mathcal{R}}_{\alpha}^p$ are bounded for each $a_p \in (0,\mathcal{A})$ (because $\underline{\epsilon}_p$ is bounded); and therefore, in this case, we can compute ellipsoidal bounds on $\tilde{\mathcal{R}}_{\alpha}^p$. This additional assumption could be enforced by adding detectors that identify anomalies in the sample mean and sample covariance of the residual. Such detectors would force the attacker to avoid arbitrarily large attack values in order to avoid detection by these additional mean and covariance detectors.

As before, we characterize, for some positive definite matrix $\tilde{\mathcal{P}}_{\alpha}^p \in \Real^{n \times n}$, outer ellipsoidal bounds of the form $\tilde{\mathcal{E}}_{\alpha}^p := \{ e_{k} \in \Real^{n} | e_{k}^T \tilde{\mathcal{P}}_{\alpha}^p e_{k} \leq 1 \}$ containing $\tilde{\mathcal{R}}_{\alpha}^p$. The results corresponding to Theorem 1, and Corollary 1 for Case 1 are stated in the following corollary.

\begin{corollary}\label{corollary3}
For given false alarm rate $\mathcal{A}$, probability $p = 1 - \mathcal{A} + a_p$, $a_p \in (0,\mathcal{A})$, threshold $\epsilon_p = \underline{\epsilon}_p = \frac{m}{\mathcal{A} - a_p} -\alpha$, and matrices $(F,L,\Sigma)$, consider the set $\tilde{\mathcal{R}}_{\alpha}^{p}$ in  \eqref{constrained_control6}. Then, for given $b \in (0,1)$, if there exists a matrix $\mathcal{P}  \in \Real^{n \times n}$ solution of the following convex optimization:
\begin{align}
\left\{ \begin{array}{ll}\label{LMI_2}
\min_{\mathcal{P}} \hspace{2mm} -\log\det \mathcal{P}, \\[1mm]
\text{s.t. } \mathcal{P}>0 \text{ and } \eqref{LMI_1},
\end{array} \right.
\end{align}
for $\bar{\omega} = \alpha + \underline{\epsilon}_p + \bar{v}_p$; then, $\tilde{\mathcal{R}}_{\alpha}^{p} \subseteq \tilde{\mathcal{E}}_{\alpha}^{p}$ (with $\tilde{\mathcal{P}}_{\alpha}^{p} = \mathcal{P}$) and $\tilde{\mathcal{E}}_{\alpha}^{p}$ has minimum volume, i.e., the $p$-probable hidden reachable set $\tilde{\mathcal{R}}_{\alpha}^{p}$ is contained in the minimum volume ellipsoid $\mathcal{\mathcal{E}}_{\alpha}^{p}=\{ e_{k} \in \Real^{n} | e_{k}^T \mathcal{P}_{\alpha}^{p} e_{k} \leq 1 \}$.
\end{corollary}

A result for redesigning the observer gain for minimizing the volume of the above ellipsoids, as in Corollary 2 for Case 1, can be stated in a similar manner as the corollary above; however, this is omitted here due to the page limit.

\begin{figure}[t]
  \centering
  \includegraphics[scale=.31]{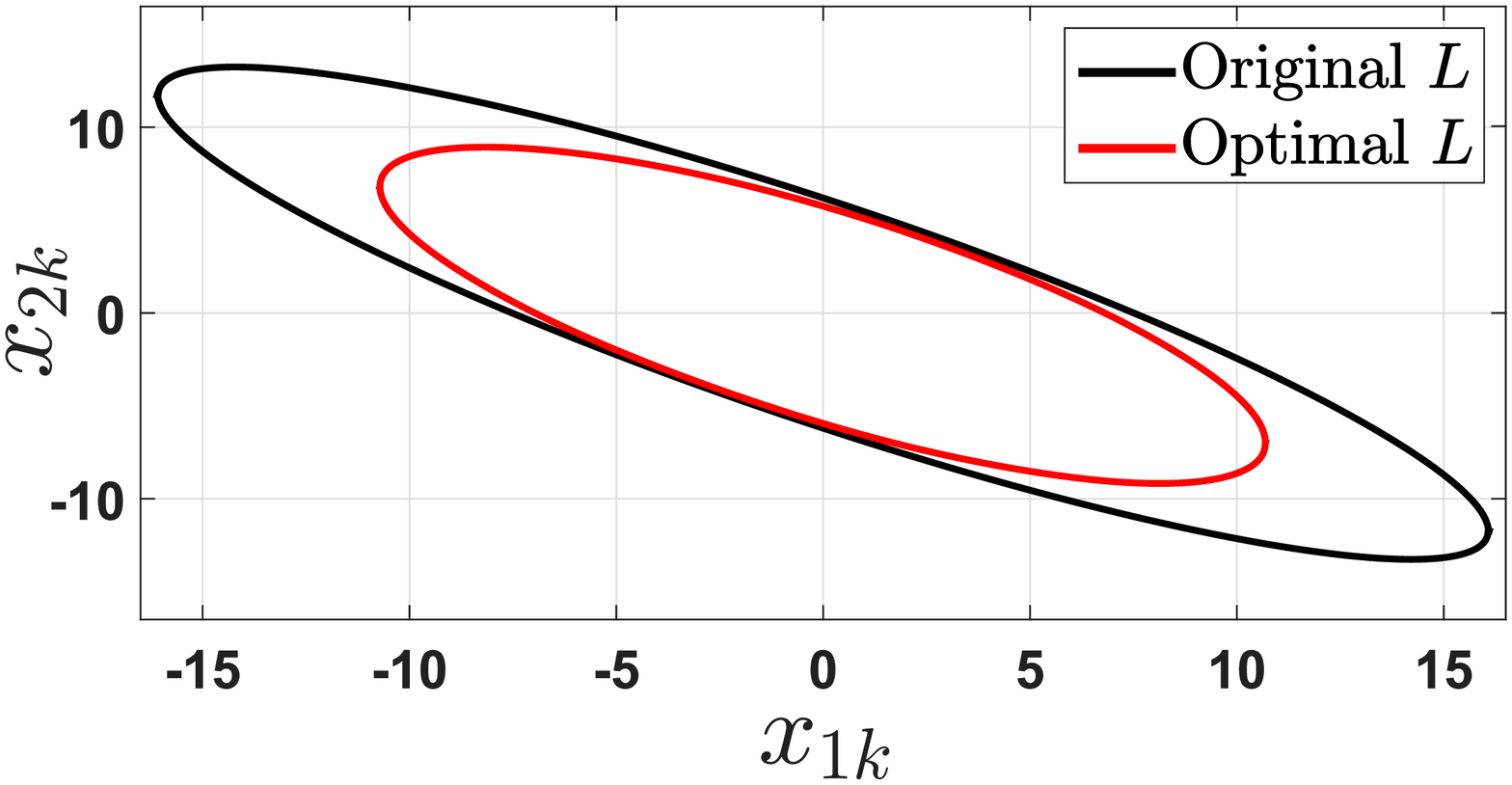}
  \caption{The improvement in the ($1-\mathcal{A}$)-probable hidden reachable set ellipsoidal bound $\mathcal{E}_{\alpha}^{1-\mathcal{A}}$, for $\mathcal{A}=0.01$, through application of Corollary 2 to design the optimal observer gain.}\label{Fig3}
\end{figure}

\begin{figure}[t]
  \centering
  \includegraphics[scale=.31]{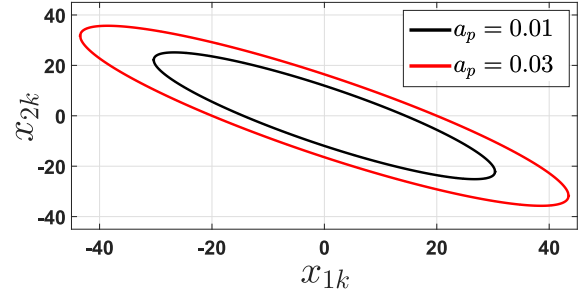}
  \caption{Ellipsoidal bound $\tilde{\mathcal{E}}_{\alpha}^{p}$ for different values of $a_p$ obtained using Corollary 3.}\label{Fig4}
\end{figure}

\section{Simulation Experiments}

Consider the closed-loop system \eqref{17}-\eqref{19} with matrices:
\begin{equation}\label{Simul}
\resizebox{.43 \textwidth}{!}
{
$
\begingroup
\renewcommand*{\arraycolsep}{1.5pt}
\left\{\begin{array}{ll}
F = \begin{pmatrix} \hspace{2.75mm}0.84 & 0.23\\-0.47 & 0.12 \end{pmatrix}, \hspace{.5mm} G = \begin{pmatrix} 0.07\\0.23 \end{pmatrix}, \hspace{.5mm}  C = \begin{pmatrix} 1&0 \end{pmatrix}, \\[4mm] \hspace{.5mm} L = \begin{pmatrix} \hspace{2.75mm} 1.16 \\-0.69 \end{pmatrix}, \hspace{.5mm} R_1 = \begin{pmatrix} \hspace{2.75mm}0.45&-0.11\\-0.11&\hspace{2.75mm}0.45 \end{pmatrix},\\[4mm] R_{0} = \begin{pmatrix} 1&0\\0&1 \end{pmatrix},\hspace{.5mm} R_2=1, \hspace{.5mm} \Sigma = 3.26.
\end{array} \right. \endgroup
$
}
\end{equation}
We start with Case 1. Using Proposition 2, the observer gain $L$ is designed such that the $H_{\infty}$ gain from the noise to the residual $r_k$ of \eqref{26} is less than or equal to $\gamma=1.86$ in the attack-free case. Consider the false alarm rates $\mathcal{A} = \{0.01,0.05,0.10,0.20\}$ and the corresponding $\alpha = \{6.63,3.84,2.70,1.64\}$, obtained using Proposition 1. The thresholds $\bar{v}_{1-\mathcal{A}}$ in (\ref{restriction}) are computed such that $\text{pr}[\norm{v_k}^2 \leq \bar{v}_{1-\mathcal{A}}] = 1-\mathcal{A}$. Because the entries on the diagonal of $R_1$ are equal and $v_k \sim \mathcal{N}(\mathbf{0},R_1)$, the random sequence $\norm{v_k}^2$, $k \in \Nat$ follows a gamma distribution, $\Gamma(\kappa,\theta)$, with shape parameter $\kappa=1$ and scale parameter $\theta=0.90$, see \cite{Ross}. It follows that, for these $\mathcal{A}$, $\bar{v}_{1-\mathcal{A}}= \{4.14,2.69,2.07,1.44\}$. For these values of $\bar{v}_{1-\mathcal{A}}$ and $\alpha$, in Figure 2, we depict the ellipsoidal bounds $\mathcal{E}_{\alpha}^{1-\mathcal{A}}$ on the $(1-\mathcal{A})$-probable hidden reachable sets $\mathcal{R}_{\alpha}^{1-\mathcal{A}}$ obtained using Theorem 1 and Corollary 1. Next, for $\mathcal{A}=0.01$, using Corollary 2, we redesign the observer gain $L$ to minimize the volume of $\mathcal{E}_{\alpha}^{1-\mathcal{A}}$ while maintaining the $H_{\infty}$ performance below $\gamma=1.86$. The obtained optimal ellipsoidal bound, $\mathcal{E}_{\alpha}^{1-\mathcal{A}}$, is depicted in Figure 3 for the optimal observer gain $L=(0.1272,-0.0160)^T$. For Case 2, let $\mathcal{A}=0.05$, $p=1-\mathcal{A}+a_p$, $a_p=\{0.01,0.03\}$, and $L$ as in \eqref{Simul}; then, the corresponding $\bar{v}_{p}$ are $\bar{v}_{p}= \{2.8970,3.5208\}$ and the $\underline{\epsilon}_p$, computed through (\ref{lower_bound}), are given by $\underline{\epsilon}_p = \{21.16,46.16\}$. In Figure 4, we show the ellipsoidal bounds $\tilde{\mathcal{E}}_{\alpha}^{p}$ on the reachable sets $\tilde{\mathcal{R}}_{\alpha}^{p}$ obtained using Corollary 3.

\begin{remark}
Many numerical results considering hidden attacks with different distributions are presented in the accompanying paper \emph{\cite{RuthsACC_Hidden_Empirical_Dist}} (Section 4). Also, extensive Monte-Carlo simulations showing the tightness of the bounds presented here are given in \emph{\cite{RuthsACC_Hidden_Empirical_Dist}}.
\end{remark}

%

\section{Conclusion}

In this paper, for a class of discrete-time LTI systems subject to sensor/actuator noise, we have provided tools for \emph{quantifying} and \emph{minimizing} the negative impact of sensor attacks on the estimation error dynamics performance given how the opponent accesses the dynamics (i.e., through the controller by tampering with sensor measurements). We have proposed to use the \emph{reachable set} as a measure of the impact of an attack given a chosen detection method. For given system dynamics and attack detection scheme, we have derived ellipsoidal bounds on these reachable sets using LMIs. Then, we have provided synthesis tools for minimizing these bounds (minimizing thus the reachable sets) by properly redesigning the detectors.



\bibliographystyle{IEEEtran}
\bibliography{ifacconf2}

\end{document}